\documentclass{jpsj-suppl}
\usepackage{txfonts} 
\usepackage{color}
\usepackage{bm}

\title{Composite nature of $\Lambda(1405)$ from the spacial structure of $\bar{K}N$ system}

\author{Kenta \textsc{Miyahara}$^{1}$ and Tetsuo \textsc{Hyodo}$^{2}$}

\inst{$^{1}$Department of Physics, Graduate School of Science, Kyoto University, Kyoto 606-8502, Japan \\
$^{2}$Yukawa Institute for Theoretical Physics, Kyoto University, Kyoto 606-8502, Japan}

\email{miyahara.kenta.62r@st.kyoto-u.ac.jp}

\recdate{December 15, 2015}

\abst{We discuss the spacial structure of the resonance state, $\Lambda(1405)$, solving the Schr\"odinger equation with the $\bar{K}N$ local potential. The potential is constructed by paying attention to the two points, the constraint from the recent experimental data and the reliability in the complex energy region. Using the new local potential, we investigate the $\bar{K}N$ spacial structure and obtain the result indicating that $\Lambda(1405)$ is the meson-baryon molecular state, that is, the composite state.  }

\kword{$\Lambda(1405)$, $\bar{K}N$ local potential, chiral unitary approach, spacial structure, compositeness}

\begin{document}
\maketitle

Most of hadrons are interpreted to consist of three quarks (baryon) and a quark-antiquark pair (meson). Aside from the ordinary structure, the existence of the exotic hadrons, such as multiquark states and hadronic molecular states, have been studied. A candidate of the exotic hadron is $\Lambda(1405)$~\cite{Dalitz}, which is considered to be the $\bar{K}N$ molecular states~\cite{Hyodo:2011ur}. The structure of $\Lambda(1405)$ has been discussed in terms of the compositeness from observable quantities~\cite{Kamiya:2015aea} and in terms of strangeness magnetic form factor from lattice QCD~\cite{Hall:2014uca}. On the other hand, in this work, we study the structure of $\Lambda(1405)$ from the viewpoint of the $\bar{K}N$ spacial structure~\cite{Miyahara:2015bya}. 

In order to obtain the $\bar{K}N$ wave function, we first construct the $\bar{K}N$ local potential used in the Schr\"odinger equation. As a base of the potential construction, we briefly introduce the chiral unitary approach~\cite{Ikeda}. 
In chiral unitary approach, the $s$-wave meson-baryon scattering amplitude is calculated by resumming the interaction terms $V_{ij}$ from chiral perturbation theory, where $i$ denote the meson-baryon channel indices.
The free parameters to fit the experimental data are the low energy constants in $V_{ij}$ and the subtraction constants in meosn-baryon loop function. In the work of Refs.~\cite{Ikeda}, these parameters are fitted to reproduce the recent precise experimental data by the SIDDHARTA Collaboration~\cite{Bazzi}, which leads to the significant reduction of the uncertainty of the $\bar{K}N$ amplitude.

Based on chiral unitary approach, we construct the local potential useful to obtain the spacial structure, following Ref.~\cite{Hyodo:2007jq}. Using the $\bar{K}N$ single-channel interaction $V_{11}^{\rm eff}$ extracted by the Feshbach projection of $V_{ij}$, the $\bar{K}N$ local potential $U$ can be constructed as
\begin{align}
U(r,E) &= \frac{e^{-r^2/b^2}}{\pi^{3/2}b^3}\frac{M_N}{2(E+M_N+m_K)}\frac{\omega_K+E_N}{\omega_K E_N}\left[V^{\rm eff}_{11}(E+M_N+m_K)+\Delta V(E) \right].  \label{eq:equivpot1}
\end{align}
$E$, $E_N$, and $\omega_K$ represent the nonrelativistic meson-baryon total energy, the nucleon energy, and the anti-kaon energy. $M_N$ and $m_K$ represent the nucleon mass and the kaon mass. Solving the Sch\"odinger equation with this potential, we can obtain the scattering amplitude from the behavior of the wave function at the sufficiently large distance. Comparing this amplitude and the original amplitude from chiral unitary approach, we determined the range parameter $b$ and the correction $\Delta V$. 
As a characteristic point of this work, we reproduce the original amplitude even in the complex energy plane, because the resonance pole is in the complex energy plane and the spacial structure of the resonance is sensitive to the pole energy. This potential can be applied to the few-body calculation like $\bar{K}NN$ system as well as the estimation of the $\Lambda(1405)$ spacial structure.

\begin{figure}[tbh]
\begin{center}
\includegraphics[width=6.5cm]{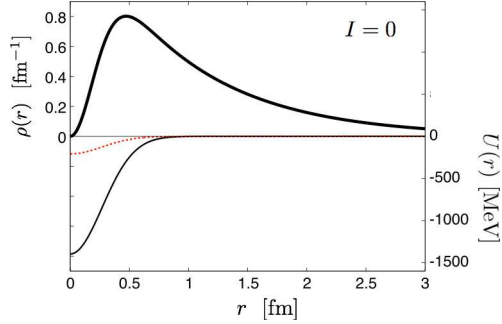}
\caption{The $\bar{K}N$ density distribution (thick line), the real part of the local potential (thin solid line), and the imaginary part of the local potential (dotted line).}
\label{f1}
\end{center}
\end{figure}

With the new local potential, we calculate the $\bar{K}N$ distance in the $\Lambda(1405)$ energy. Solving the Schr\"odinger equation, we obtain the $\bar{K}N$ wave function. The result of the density distribution and the potential behavior are shown in Fig.~\ref{f1}.
In order to study the structure more quantitatively, we calculate the root mean square distance $\sqrt{\langle r^2 \rangle}$. As explained in Ref.~\cite{Miyahara:2015bya}, for the complex and energy-dependent potential, the expectation value is calculated as
\begin{align}
\langle r^2 \rangle_G \equiv {\int d{\bm r}\ r^2\left( 1-\frac{\partial U(r,E)}{\partial E} \right)\psi^2} \Big{/}
{\int d{\bm r}\ \left( 1-\frac{\partial U(r,(E)}{\partial E} \right)\psi^2}.  \label{eq:EdepGamowexp}
\end{align}
The result is $\sqrt{\langle r^2 \rangle_G} = 1.04-0.61i\ {\rm fm}$. For the weakly binding case, that is, the large distance case, the mean distance can be estimated only from the eigen energy $E$ and the reduced mass $\mu$,
$\sqrt{\langle r^2 \rangle} \sim {1}/{(2\sqrt{-\mu E})}=0.85-0.58i\ {\rm fm}$.
Because this value is similar to the distance from the wave function, we can suppose that the $\bar{K}N$ system is largely extended. Though we can gain some insight from Eq.~\eqref{eq:EdepGamowexp}, it is difficult to interpret the complex radius. As shown in Ref.~\cite{Miyahara:2015bya}, because the dumping of the wave function outside the potential can be extracted from
\begin{align}
\langle r^2 \rangle \equiv {\int d{\bm r}\ r^2 |\psi(r)|^2}\Big{/}{\int d{\bm r}\ |\psi(r)|^2}.  \label{eq:usualexp}
\end{align}
The derivative of the potential as in Eq.~\eqref{eq:EdepGamowexp} can be neglected because the effect of this term is found to be small. With Eq.~\eqref{eq:usualexp}, the result is $\sqrt{\langle r^2 \rangle} = 1.44\ {\rm fm}$. Considering the charge radii of the proton ($\sim$ 0.85 fm) and the kaon ($\sim$ 0.55 fm), we find that the $\bar{K}N$ distance is relatively larger. This result supports the $\bar{K}N$ molecular state picture of $\Lambda(1405)$ rather than the elementary state picture like the three quark state, in accordance with Refs.~\cite{Kamiya:2015aea,Hall:2014uca}.



\end{document}